\documentclass[aps,prb,twocolumn,epsfig,showpacs]{revtex4}
\usepackage{amssymb,amsbsy,amsmath,bm,epsfig,hyperref,verbatim}
\graphicspath{
  {./figs/}
}

\begin{document}
\title{Orbital Mott transition in two dimensional Pyrochlore lattice}
%\shorttitle{} %Insert here a short version of the title if it exceeds 70 characters

\author{Abhinav Saket$^1$ and Rajarshi Tiwari$^2$}

\affiliation{
  % $^1$Harish-Chandra  Research Institute, Chhatnag Road, Jhusi, Allahabad 211019, India\\
  $^1$Samastipur College, Samastipur, Bihar-848134, India\\
  %Harish-Chandra  Research Institute, Chhatnag Road, Jhusi, Allahabad 211019, India\\
  $^2$School of Physics and CRANN, Trinity College, Dublin 2, Ireland
}

%\shortauthor{Tiwari \etal}
%\institute{
%  \inst{1} Harish Chandra Research Institute, Chhatnag Road, Jhunsi, Allahabad-211019, India\\
%  \inst{2} School of Physics and CRANN, Trinity College, Dublin 2, Ireland
%}
%\pacs{71.30.+h}
%{Metal-insulator transitions and other electronic transitions}
%\pacs{71.27.+a}
% {Strongly correlated electron systems; heavy fermions}
%\pacs{74.25.Gz}{Optical properties}
%\pacs{74.70.Kn}{Organic superconductors}
%\keywords{triangular lattice, Hubbard model, frustration, Mott transition}
%\email{rajarshi@hri.res.in, abhinav.saket.sinha@gmail.com}

\begin{abstract}
  We study orbital Mott transition in two dimensional pyrochlore lattice,
  using a two orbital Hubbard model with only inter-orbital electronic hopping.
  We use a real space Monte Carlo based approach to study the model
  at finite temperature, and establish temperature-interaction
  phase diagrams that highlight the Mott transition, orbital
  ordering, and spectral trends, and possible window of pseudogap.
  Due to only inter-orbital hopping, the Mott insulator `generates`
  ferro exchange resulting in ferro-orbital ordering, with
  T$_{corr}/t$ peaked at $\approx 0.2$ around $U/t \approx 6$.
  The optical conductivity shows unusual two peak feature
  due to two dimensional pyrochlore lattice.
\end{abstract}

\date{September 3, 2019} % {\today}

\maketitle

\section{Introduction}
The Mott transition is one of the most widely discussed phenomenon in strongly correlated systems\cite{mott-orig-1, mott-orig-2}.
It manifests itself at `integer' filling when electrons localise due to electron-electron
interaction, resulting in an insulating state, called Mott insulator. The `Mott problem',
i.e., understanding why the transition occurs and its consequences, is most commonly
studied through single band Hubbard model\cite{mott-rev-tokura} on different lattices.
The addition of orbital degrees of freedom opens more exciting possibilities, such
as interplay of orbital and spin degrees of freedom\cite{orb-phys-khomskii}, orbital
selective Mott transition\cite{orb-sel-mott}, where some orbitals localise, while
other remain itinerant.
These possibilities, are, however, explored at the cost of more complex models
that include intra and inter orbital repulsions and exchange energies\cite{mult-orb-model}.
Formally, each energy scale/parameter adds a dimension to the parameter space,
making it harder to comprehend. As a result, studies geared towards model solving
usually reduce the parameter space, by assuming simpler electronic hopping
structure, and choosing reduced set of interactions, such as absence of
inter-orbital repulsion\cite{florens}, fixing interaction to some representative
values\cite{tomoko, becca}, or choosing same intra and inter-orbital repulsions
by removing Hund's coupling\cite{grandi}.

\begin{figure}[t]
\centering
\psfig{file=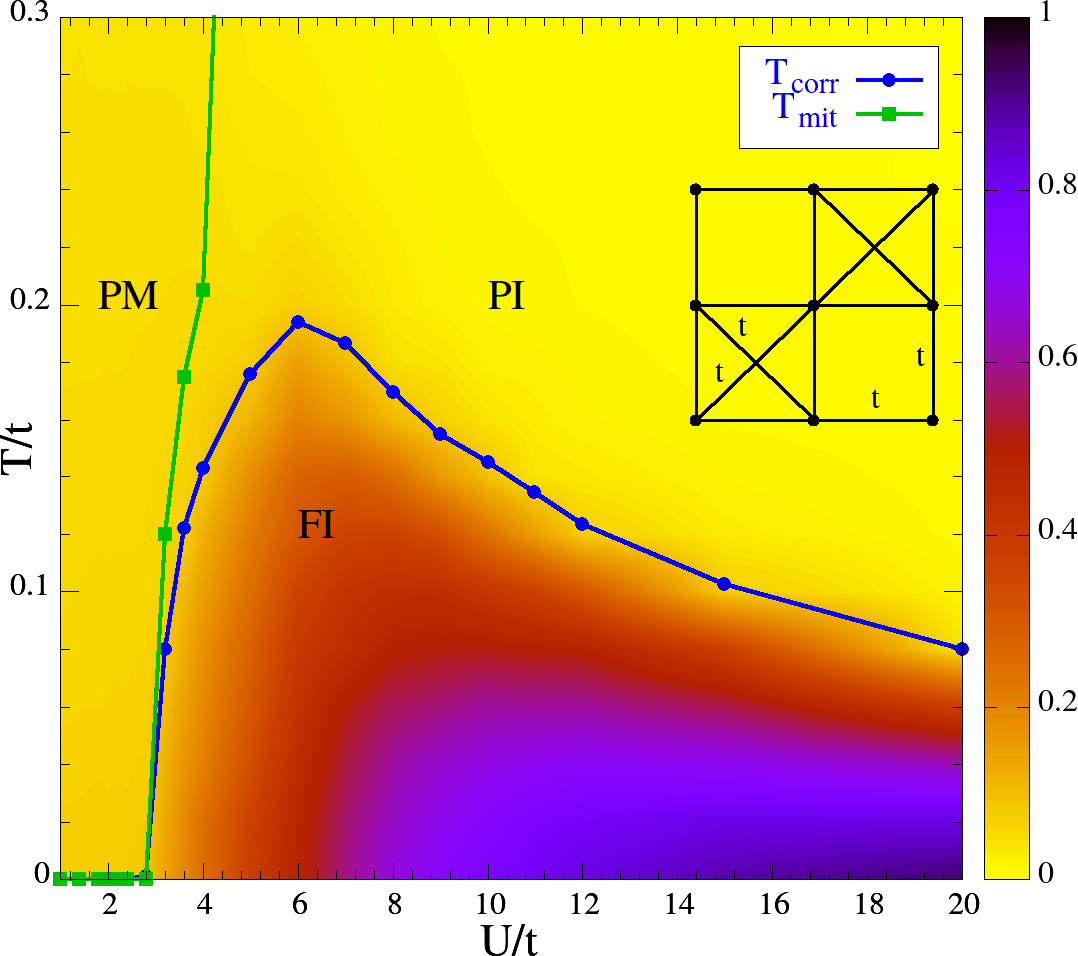,width=8cm}
\caption{Colour online: The phase diagram of the Furukawa model on checkerboard lattice
  (shown in the inset) in the $U/t, T/t$ plane.
  The color map denotes the value of maxima of S(${\bf q}$) at given temperature and $U/t$. 
  The maxima in this case corresponds to ${\bf q}=(0,0)$, which gives the ferromagnetic order.
  The blue (circle) curve shows $T_{corr}$ temperature, at which magnetic correlation grow up
  to lattice size. The green (square) curve shows the metal-insulator transition, determined
  by the sign of $d\rho/dT$. (See text).
}
\label{fig-phd-sf}
\end{figure}

%One example of orbital dependent physics is the orbital-selective Mott transition, where some of
%the orbitals become localised by electron correlations, while the others still remain itinerant.
To solve the Hubbard model for finite temperature, DMFT has been a
method of choice, be it single orbital model\cite{georges-rmp, kotliar-prl},
or multi-orbital model\cite{kotliar, florens, tomoko}.
Many qualitative features of orbital-degenerate Mott transition are
found to be available in single band model as well\cite{kotliar},
however new features emerge, that require non-trivial treatment
of multiple orbitals, such as different scaling of the critical couplings
$U_{c1}$ and $U_{c2}$ as function of the number of orbitals\cite{florens}.
Recently, Kawakami et al studied the Mott transition in the three-orbital
Hubbard model and investigated how the orbital level splitting affects the
Mott transition in the case of two electrons per site using (DMFT) combined
with continuous-time quantum Monte Carlo simulations\cite{tomoko}.
A general mechanism for the coexistence of both itinerant and localised
conduction electrons has been proposed\cite{hassan} and the orbital selective
Mott transition in two-band Hubbard models with different bandwidths
has been explored using another form of DMFT\cite{akihisa}.

Within DMFT, or its cluster variants\cite{maier-cdmft, sato-cdmft},
the correlated lattice system is mapped to one or more correlated
sites coupled with non-interacting bath of electrons, and
one gets to solve the quantum problem at the cost of ignoring
spatial fluctuation in the lattice.
The neglect of spatial correlations and the lack of visual intuition
about the transition, motivated us in past to explore a complementary
real-space Monte Carlo based approach (discussed later), which
provides a reasonable description of Mott physics in a real space
setting and allows a certain degree of visualization.

In this paper, we use this approach to solve a two orbital
Hubbard model, inspired from the recent interest
in pyrochlore compounds for studying the effects of
spin-orbital interplay and geometrical frustration\cite{gardner}.
The metal-insulator transition (MIT) in Mo pyrochlore oxides $(R_2Mo_2O_7)$\cite{motome4}
(where R is rare earth metal) and role of its frustrated lattice
structure has been extensively studied earlier for Ir based pyrochlore\cite{matsuhira, nakatsuji}.
The evolution of charge dynamics at metal-insulator transition
has been experimentally investigated for $Nd_2(Ir_{1-x}Rh_x)_2O_7$
where the spin-orbit interaction as well as the electron correlation
is effectively tuned by the doping level $x$\cite{ueda}.
The transition from ferromagnetic metal to spin glass insulator and
paramagnetic metal has been observed with increase of the radius of
rare earth metal ion $R^{3+}$ and external pressure due to the competing
double exchange and super exchange interactions on the frustrated
lattice\cite{iguchi1}. Likewise, the role of orbital degrees of freedom
has been debated in metal insulator transitions in various pyrochlore
oxides\cite{hanasaki}. 

A number of theoretical attempts have been made to explore the
antiferromagnetism\cite{fritsch}, frustration\cite{fujimoto, ishizuka, carter},
Hall effect\cite{disseler} etc, through a variety of models. Ground state\cite{motome4}
and finite temperature phase diagrams\cite{motome1} have been established,
showing transition from ferromagnetic to spin glass, or cooperative paramagnetic phase.
However, these transitions are seen primarily in terms of double-exchange model without
electron-electron interaction, or within weak interaction limit. Furukawa
{\it et al}\cite{motome4} (Fig. 2) illustrated an elegant schematic of
phases in terms of interaction and super-exchange phase diagram, which shows
the possibility of orbital Mott transition in the ferromagnetic spin
background in the weak super-exchange limit. We wish to explore this
transition in detail.

\begin{figure}[t]
  \centering
  \psfig{file=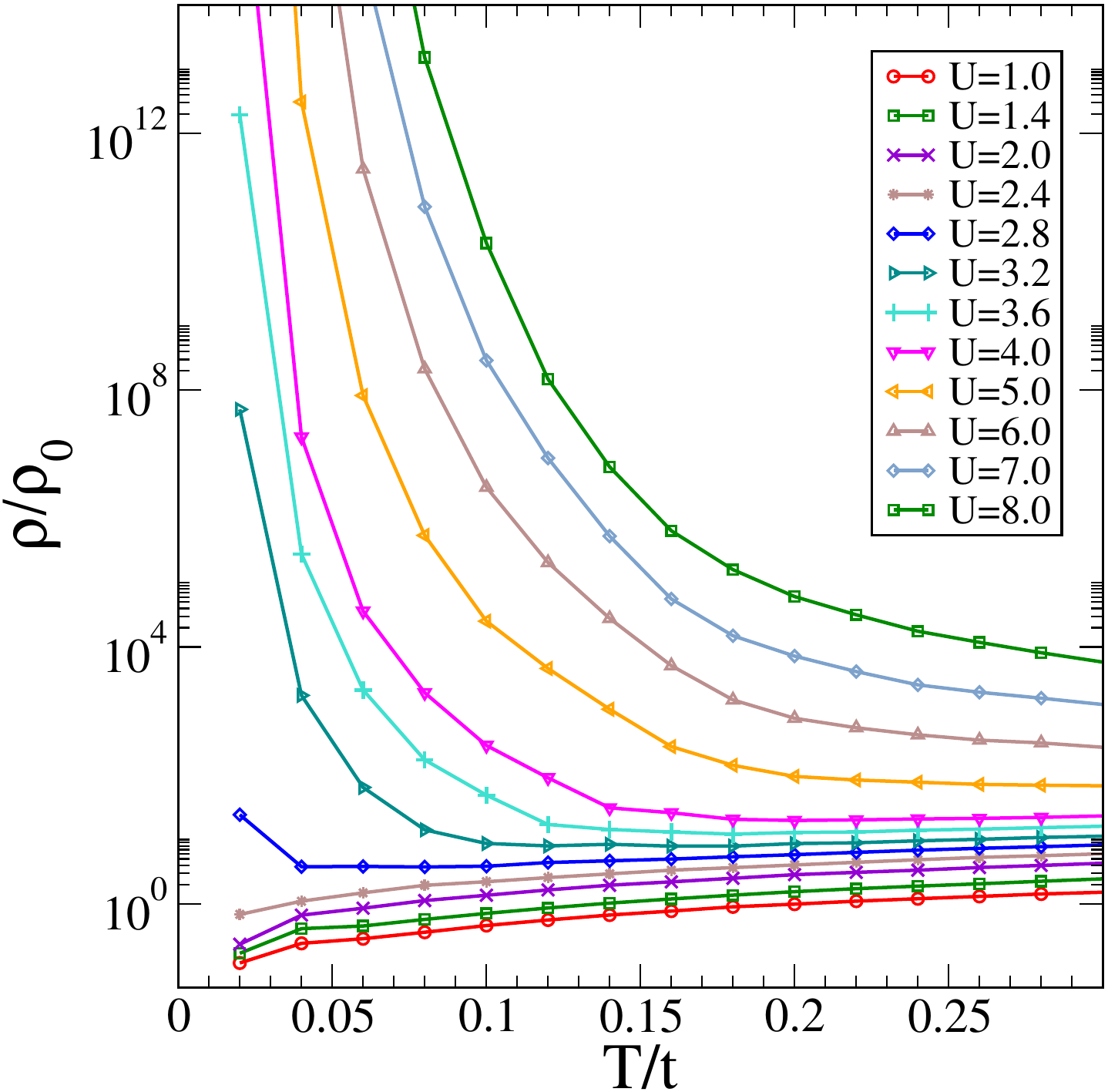,height=4.5cm}
  \psfig{file=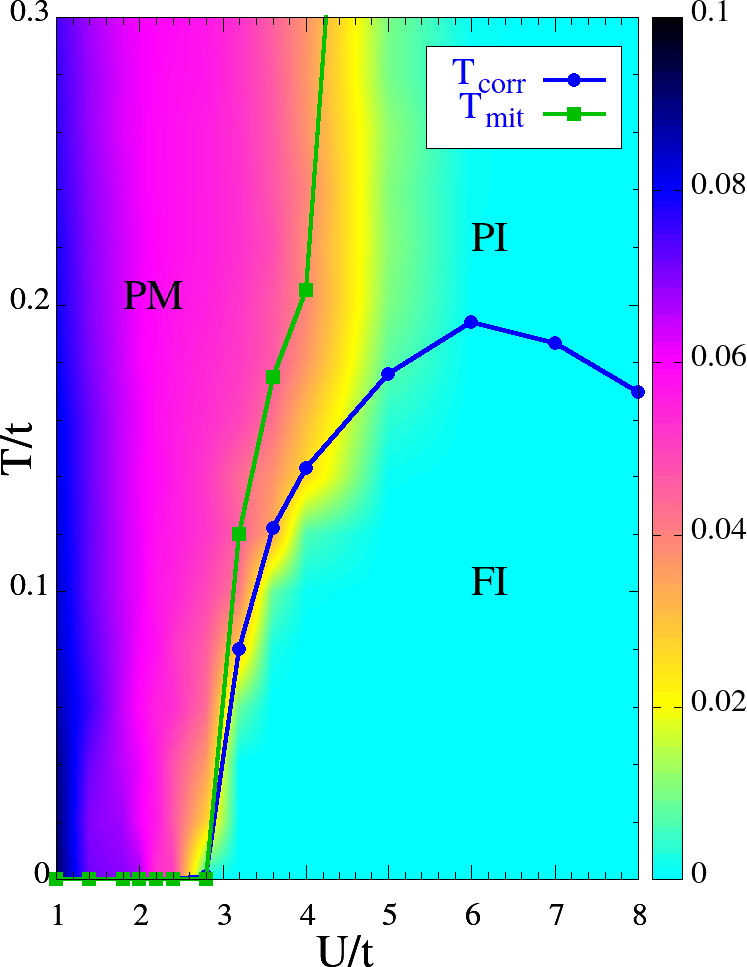,height=4.5cm}
  \caption{Colour online: (Left) Temperature dependence of the resistivity at different $U/t$.
    (Right) Colour map of the density of state at Fermi level in $U-T$ plane. The blue
    and green curves are for the magnetic transition $T_{corr}$ and insulator metal
    transition, same as in Fig.\ref{fig-phd-sf}. (See text)}
  \label{rho}
\end{figure}

\section{Model}
The lattice structure of $R_2Mo_2O_7$ is composed of two intervening pyrochlore lattices formed
by Mo cations and R cations.
The Mo cation is surrounded by octahedra of oxygens $(MoO_6)$, which splits the five fold
degenerate $d$-orbitals into three and two fold degenerate $t_{2g}$ and $e_g$ orbitals
respectively.
Further, the distortion of the $MoO_6$ octahedra along local (111) axis (towards centre
of each Mo tetrahedra) splits the $t_{2g}$ levels into lower single $a_{1g}$ level (below
Fermi level) and higher two fold degenerate $e'_g$ levels (above Fermi level)\cite{solovyev}.
Mo being $Mo^{4+}$ cation in these compounds, strong Hund's coupling results in the fully
occupied single $a_{1g}$ up-spin band, well below Fermi level, and half-filled two-fold
degenerate $e^{\prime}_g$ up-spin bands. We start with the following two band double exchange
model, previously proposed by Furukawa {\it et al}\cite{motome4} for various pyrochlore
systems to describe the electronic motion for the compound, including kinetic energy,
coulomb interaction, Hund's coupling and anti-ferromagnetic super-exchange:
\begin{eqnarray}
\label{DEmodel}
  H&=& -\sum_{\langle ij\rangle, \alpha\beta\sigma}t_{\alpha\beta}\left( c^{\dagger}_{i\alpha\sigma}c_{j\beta\sigma}+ h.c. \right)
       + J_{H}\sum_{i\alpha}\vec{S}_i\cdot\vec{s}_{i,\alpha} \nonumber\\
   &+& \sum_{i\alpha\beta\alpha'\beta'\sigma\sigma'}U_{\alpha\beta\alpha'\beta'}c^{\dagger}_{i\alpha\sigma}c^{\dagger}_{i\beta\sigma'}c_{i\beta'\sigma'}c_{i\alpha'\sigma}\nonumber\\
  &+& J_{AF}\sum_{\langle ij\rangle} \vec{S}_i\cdot\vec{S}_j
\end{eqnarray}
where, $c^{\dagger}_{i\alpha\sigma}$ creates an electron at site $i$, orbital $\alpha$ and spin $\sigma$.
$\vec{s}_{i\alpha}$ is electronic spin operator for site $i$, orbital $\alpha$. $\vec{S}_i$ is core spin
at site $i$.
The first term denotes the kinetic energy of itinerant $e'_g$ electrons with spin $\sigma$
and orbitals $\alpha$ running over two degenerate orbitals (say 1, and 2) of the $e'_g$ band.
$t_{\alpha,\beta}$ is the electronic hopping.
Second term, with coupling $J_H$, denotes double exchange coupling between itinerant $e'_g$ electrons
with localised $a_{1g}$ electron, treated as core spins.
Third term denotes the anti-ferromagnetic super-exchange among localised $a_{1g}$ electrons with strength $J_{AF}$.
$J_{AF}=$ is approximately set by $t_{a1g}^2/U_{a1g}$ where $t_{a1g}$ is the transfer integral between the
$a_{1g}$ orbitals and $U_{a1g}$ the intra-orbital Coulomb repulsion in the $a_{1g}$ orbital.
The last term denotes on-site coulomb interactions between $e'_g$ including intra and inter-orbital repulsions.

\begin{figure*}[htp]
  \centering
  \psfig{file=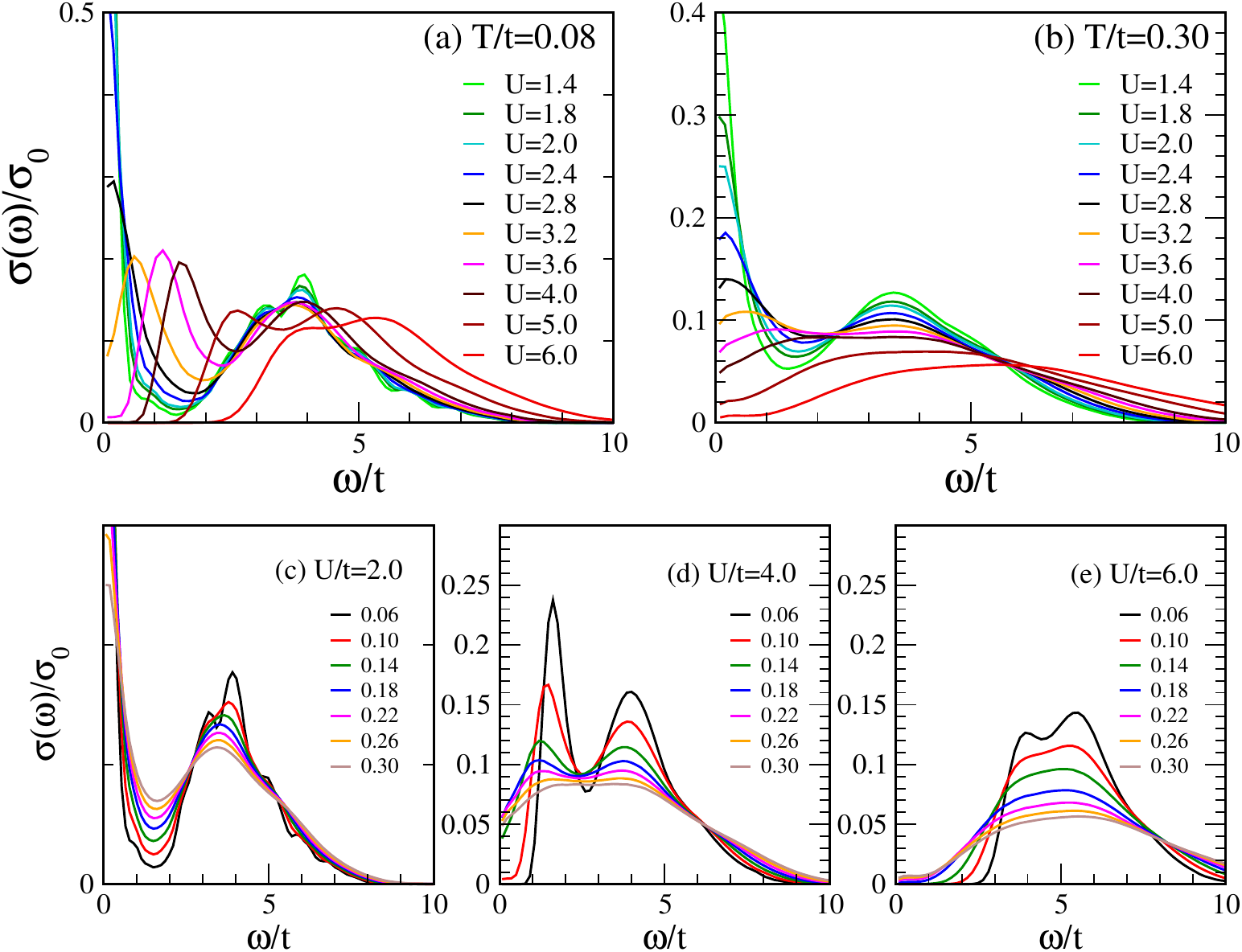,width=8.5cm, height=6.1cm}\hspace{0.3cm}
  \psfig{file=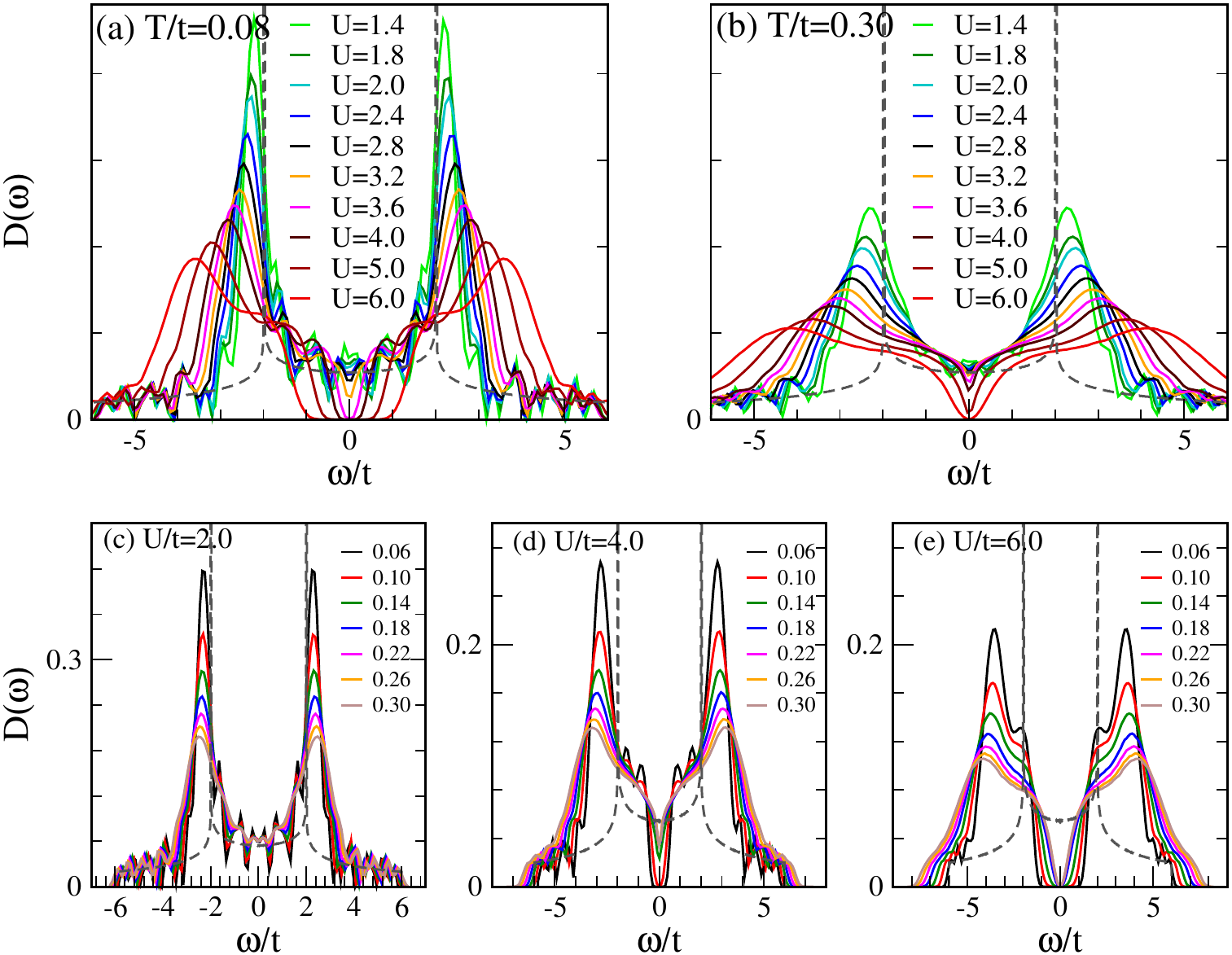,width=8.5cm, height=6.05cm}
  \caption{Colour online:
    The left panel (a)-(e) shows optical conductivity $\sigma(\omega)$,
    and the right panel (a)-(e) shows the DOS. In both the panels,
    we show the $U/t$ dependence at low (a) T/t=0.08, and high (b) T/t=0.30 temperature,
    while we show $T/t$ dependence at three representative strengths of interaction
    (c) $U/t = 2$, (d) $U/t = 4$, (e) $U/t = 6$. (see text)
  }
  \label{opt-dos}
\end{figure*}

We study the above two band double exchange (DE) model (\ref{DEmodel}) on the two dimensional pyrochlore
lattice, which is essentially `checkerboard lattice' (shown in the inset of Fig.\ref{fig-phd-sf}),
in the limit $J_{AF}=0$. This gets further simplified in the $J_{H}\rightarrow \infty$ limit as follow.
Rotating the axis of quantization of every fermionic operator $c_{i,\alpha}$ from
universal z-axis to the direction of the core spin $S_i(\theta_i,\phi_i)$ at every site,
\begin{eqnarray}\nonumber
  \left[\begin{array}{c}c_{i,\uparrow}\\c_{i,\downarrow}\end{array}\right]=U(\theta_i, \phi_i)\left[\begin{array}{c}p_i\\a_i\end{array}\right]
  \mbox{,\quad where~~}\\
  U(\theta_i,\phi_i)=exp\big(-i\frac{\phi_i}{2}\sigma^z\big)exp\big(-i\frac{\theta_i}{2}\sigma^y\big).
\end{eqnarray}
renders the Hund's term diagonal in spin, and in $J_{H}\rightarrow \infty$ limit,
the parallel $p_i$ state gets projected out from the Hamiltonian, as have higher energy.
By shifting the zero of the energy by $J_{H}$, we get the following low energy Hamiltonian,
written in terms of the $a_i$ operators with their axes aligned opposite to the
localised spin at each site \cite{motome4}.
\begin{eqnarray}
  \label{mainmodel}
  H=-\sum_{\langle ij\rangle\alpha\beta}t_{i\alpha,j\beta}\left( a^{\dagger}_{i\alpha}a_{j\beta}+ h.c. \right)+ U\sum_{i,\alpha\neq\beta } n_{i\alpha}n_{i\beta} 
\end{eqnarray}
where $a_{i\alpha}$ is spinless fermion operator and $n_{i\alpha}= a^{\dagger}_{i\alpha}a_{i\alpha}$.
$U$ is inter-orbital interaction and the electronic hopping element $t_{i\alpha, j\beta}$ becomes
angle dependent, $(\theta_i,\phi_i)$ being the angles of spin $\vec{S_i}$.
\begin{eqnarray}
  \label{hopelement}
  t_{i\alpha, j\beta}=t_{\alpha\beta}(\cos{\frac{\theta_i}{2}}\cos{\frac{\theta_j}{2}}
+e^{i(\phi_i-\phi_j)}\sin{\frac{\theta_i}{2}}\sin{\frac{\theta_j}{2}} )
\end{eqnarray}

Its easy to see from eq.(\ref{hopelement}) that $t_{i\alpha, j\beta} = t_{\alpha,\beta}$
if the spins are parallel ($\theta_i = \theta_j, \phi_i = \phi_j$),
and $t_{i\alpha, j\beta} = 0$ when they are anti-parallel ($\theta_i = \pi-\theta_j, \phi_i = \pi+\phi_j$).
Thus, for $t_{i, j}$ to be maximum, nearest neighbouring spins have to be parallel to each other.
In general, a non-trivial core spin state would generate complex spatial texture of hopping
for electrons to delocalise.
However, since we consider $J_{AF}=0$, the ground state for the core spins
is ferromagnetic(FM), with its $T_c$ driven by the kinetic energy, which we
assume to be large compared to $\sim t_{ij}^2/U$, so we can approximate the
the core spins to be frozen in FM state, for which the hopping becomes site
independent $t_{i\alpha,j\beta} = t_{\alpha\beta}$. We comment more on this
later during the discussion.
%In parallel spin configuration, the hopping amplitude $t_{i, j}$ becomes site and spin independent.

Because of anisotropy of the $e^\prime_g$ orbitals and relative angle of Mo-O-Mo bond \cite{motome4,ichikawa},
the relative strength of $t_{\alpha=\beta}$ and $t_{\alpha\neq\beta}$ can be expressed as\cite{motome4}
$\frac{t_{\alpha\neq\beta}}{t_{\alpha=\beta}}=\frac{3-\cos\delta}{3+\cos\delta}$.
In Mo pyrochlore oxides $\delta>90^0$, and in Mo based systems about 130$^0$\cite{moritomo}
so the inter orbital hopping $t_{\alpha \ne \beta}$ is larger than the intra orbital
hopping $t_{\alpha = \beta}$ \cite{motome4}. So we choose,
for simplicity, $t_{\alpha = \beta} = 0$, and set $t_{\alpha \ne \beta}=t$. We will comment
on the inclusion of the intra-orbital hopping in the section \ref{discuss}

Treating the two orbital as
`pseudo-spins' $\uparrow, \downarrow$, we get the following `orbital-Hubbard model':
\begin{eqnarray}
\label{furukawa}
  H= -t\sum_{\langle ij\rangle}\left( a^{\dagger}_{i\uparrow}a_{j\downarrow}+a^{\dagger}_{i\downarrow}a_{j\uparrow} + h.c. \right)+ U\sum_i n_{i\uparrow}n_{i\downarrow}
\end{eqnarray}

Notice that, in this model, electrons delocalise through hopping
alternatively via the two `pseudo-spin' channels, which, as opposed
to the usual Hubbard model, would generate `ferromagnetic' interaction
among the neighbouring pseudo-spins in large $U/t$ Mott state, because
due to the presence of `alternate hopping', an electron of spin $\uparrow$
at site $i$ can virtually hop to neighbouring site $j$ and back, provided the
electron at site $j$ is in $\downarrow$ state, generating $\sim \frac{t^2}{U}$
exchange. From now on what we refer to as `magnetic' is in context of pseudo-spins,
and hence should be considered appropriately as orbital version of magnetism,
for example, ferromagnetism is really a ferro-orbital order.

We use static auxiliary field (SAF) approximation\cite{tiwari1,tiwari2,nyaya1}, earlier
applied to Hubbard model on different lattices to the solve this model in
real space.
We use Hubbard Stratonovich transformation\cite{schulz, hirsch} in
terms of a vector field ${\bf m}_i(\tau)$ and a scalar field $\phi_i(\tau)$ at
each site to decouple the $U n_{i\uparrow}n_{i\downarrow}$ interaction,
retaining rotation invariance of the Hubbard  model. We treat the ${\bf m}_i$
and $\phi_i$ as classical fields, i.e., neglect their time dependence, but
completely retain the spatial fluctuations in ${\bf m}_i$, while we treat
$\phi_i$ at saddle point level, i.e.,
$\langle \phi_i\rangle =\frac{U}{2}\langle n_i \rangle = \frac{U}{2}$ (at half filling). 
We have used this approach successfully in past for Mott transition on
anisotropic triangular lattice\cite{tiwari1}, FCC\cite{tiwari2} and pyrochlore\cite{nyaya1}.
Within this approximation, the model (eq.\ref{furukawa}) maps to the following
\begin{equation}
  \label{eq:hub-hs}
  H_{eff} = -t\sum_{\langle ij\rangle\alpha}a^{\dagger}_{i\alpha}a_{j-\alpha} -\frac{U}{2}\sum_i {\bf m}_i\cdot\vec{\sigma}_i + \frac U4 \sum_i {\bf m}_i^2
\end{equation}
which describes the motion of electrons coupled classical auxiliary fields ${\bf m}_i$.
The ground state of \ref{eq:hub-hs} is given by $\{{\bf m}_i\}$ that minimizes
the total energy. The thermal physics is accessed using Monte Carlo (MC) sampling
of the auxiliary field $\{{\bf m}_i\}$ that have distribution
$P(\{{\bf m}_i\}) \propto \textrm{Tr}_{p,p^{\dagger}}e^{-\beta H_{eff}}$.
We use single site update scheme, where we attempt an update ${\bf m}_i\rightarrow {\bf m'}_i$
at site ${\bf X}_i$. We compute the energy cost of the attempted update $\Delta E = E\{{\bf m}_i\}-E\{{\bf m'}_i\}$
by numerically diagonalizing the electronic Hamiltonian, and use Metropolis algorithm to update the auxiliary field.
To access large lattices within limited time, we use travelling cluster
algorithm\cite{tca} for estimating the update cost of MC, where instead
of diagonalizing the the full lattice, we calculate the energy cost of
and update ${\bf m}_i$ by diagonalizing a cluster of smaller, fixed size
defined with auxiliary field around the reference site. We have extensively benchmarked
this cluster scheme\cite{tca}. %We will discuss the limitations of the method later in the paper.
The results we show in the next section, are averaged over equilibrium MC configurations.

\section{Results}
Most of our results are based on MC done on $N=24\times 24$ lattice, with clusters of size $N_c = 8\times 8$,
which is big enough considering finite size effects, and computational resources.
We annealed the system from high temperature $T/t \approx 0.3$ for different values of $U/t$.
We probe the magnetic correlation and transition temperature $T_{corr}$ through thermal average of
the structure factor defined as
$S(q) = \frac{1}{N^2} \sum_{ij} \langle {\bf m}_i\cdot{\bf m}_j\rangle e^{ i{\bf q}\cdot({\bf R}_i-{\bf R}_j)}$
at each temperature. Its rapid growth at few specific ${\bf q}$ at some critical temperature
serves us as the onset of a transition to a state with long range order, giving us an
estimate of $T_{corr}$. Throughout the $U$ window, the maxima of the structure factor
occurs at ${\bf q}=(0,0)$, which describes `ferromagnetic' order of the pseudo-spins.

The conductivity of the two dimensional system is first
calculated as follows (ref.\cite{allen}), using the Kubo formula:
\begin{eqnarray}
\sigma^{x}(\omega) &=& \frac{\sigma_{0}}{N}\sum_{\alpha,\beta}
{ {n_{\alpha}-n_{\beta}} \over {\epsilon_{\beta\alpha}} } 
|\langle \alpha|J_x|\beta\rangle|^2
  \delta(\omega-\epsilon_{\beta\alpha})
\\
J_x &=& -it\sum_{i,\vec{\delta},\sigma}\left[\vec{\delta}\cdot \hat{x}~a^{\dagger}_{i,\sigma}a_{i+\vec{\delta},\sigma}-\textrm{hc}\right]
\end{eqnarray}
Where, $\epsilon_{\beta\alpha} = \epsilon_{\beta}-\epsilon_{\alpha}$, $J_x$ is current operator,
and $\vec{\delta}$ runs over the set of vectors connecting the neighbouring sites.
The d.c. conductivity is the $\omega \rightarrow 0$ limit of 
the above result.
$\sigma_{0}$=$\frac{\pi e^2}{\hbar}$, the scale for
two dimensional conductivity, has the dimension of {\it
conductance}.
$n_{\alpha}=f(\epsilon_{\alpha})$ is the Fermi function,
and $\epsilon_{\alpha}$ and $|\alpha\rangle$ are 
respectively the single particle eigenvalues and eigenstates of
$H_{eff}$ in a given background \{${\bf m}_i$\}. The thermal average
of the conductivity that we show later is averaged over equilibrium
\{${\bf m}_i$\} configurations generated through MC, i.e.
$\sigma(\omega, T) = \langle\langle \sigma^x(\omega)\rangle\rangle_{MC}$ .

We first summarize our results in the $U-T$ phase diagram, shown in Fig.\ref{fig-phd-sf},
where the color map represents the value of the structure factor at ${\bf q}=(0,0)$, indicative of
ferromagnetic order. The blue curve shows the $T_{corr}$ as function of $U/t$.
We see that there is a critical $U_c/t\approx 3$ at $T/t=0$, separating
ferromagnetic insulator (FI) with an orbital-insensitive metal.
Before $U_c$, we have no long range order. The $T_{corr}$ starts to increase after
$U_c$ up to $U/t=6$, after which it decreases monotonically (at large $U/t$ it goes
as $\sim t^2/U$).
The green curve defines metal-insulator boundary $U_c(T)$, based on change of sign of
resistivity derivative $d\rho/dT$.

The phases are as follows. For $U < U_c$ we have `paramagnetic' metal (PM), characterized
by increasing resistivity with temperature, and no long range spatial correlation.
For $U > U_c$, we have paramagnetic insulator (PI) state at high temperature, and
ferromagnetic insulator (FI) at low temperature.
Both are characterized by decreasing resistivity with temperature.
The T$_{corr}$ curve (shown in blue circles in Fig.\ref{fig-phd-sf}) separates FI
state with ferromagnetic order with the PI state that has no long range order.

%\subsection{Resistivity}
In left panel of Fig.\ref{rho}, we show the temperature dependence of the resistivity $\rho(T)$
at several $U/t$, which neatly demonstrates the MIT. At low $U/t$ the resistivity is metallic,
and increasing $U/t$ results in progressively higher, yet metallic resistivity up to critical
interaction strength $U_c/t\sim 3$, after which we have insulating resistivity, increasing
with $U/t$.

In the metallic window of $U/t$, the resistivity decreases relatively slowly when T/t is lowered,
while in the insulating window, the change is rather drastic due to presence of the Mott gap.
The right panel of the Fig.\ref{rho} re-highlights the phase diagram in terms of
the density of state (DOS) at Fermi level ($D_{Fermi}$) shown as color map.
Clearly, the insulating PI and FI phases show absence of the DOS at Fermi level, while
deep in the metallic side we have non-trivial $D_{Fermi}$, which are close to tight-binding (TB)
$D_{Fermi}$ at low, or rather zero Temperature. However, in the metallic side close to MIT,
the $D_{Fermi}$ decreases with increasing temperature. This occurs due
to thermally generated auxiliary fields $\{{\bf m}_i\}$, which grow larger with
temperature. Close to MIT line, the `locus' of constant $D_{Fermi}$ seems to follow
the MIT line, with $D_{Fermi}\ge 20\%$, of its maxima, showing that the system
becomes insulating before the $D_{Fermi}$ gets depleted, or the Mott gap opens.
Next, we discuss the optics and DOS.

Close to the MIT boundary on the insulating side, $U/t \approx 3.2$, the
resistivity has a weak non-monotonic behaviour (Fig.\ref{rho} right panel).
The ground state at U=3.2, being close to the MIT, but on insulating side
has a small gap. As the T is increased, angular fluctuations of $m_i$ weaken
the long-range order, which lowers this gap, reducing the resistivity till the
gap closes. Further increase to rather large T, the magnitude fluctuations of
$m_i$ become large, which push the DOS slowly away from Fermi level. This we
think results in slow increase of resistivity, and this non-monotonic behaviour.
This behaviour is seen close to MIT boundary, when the gap is small.
The MIT boundary for Mott transition is often non-monotonic, though
why it is more prominent some system than others is poorly understood.

%\subsection{Optical conductivity and Density of state}
In Fig.\ref{opt-dos} we show optical conductivity (left panel), along with the DOS (right panel).
Figs.\ref{opt-dos}(a)-(b), show the $U/t$ dependence of the optical conductivity (left panel)
and DOS (right panel) at low $T/t=0.08$, and high $T/t=0.3$ temperatures.
Similarly, Figs.\ref{opt-dos}(c)-(e) show $T/t$ dependence of of the optical conductivity and DOS
at three representative values of interaction strength (c) $U=2$ which is metallic, (d) $U=4$
which is insulator close to MIT, and (e) $U=6$ deeper in insulating side.
In the right panel for DOS, we also show the TB DOS in dotted line for reference.

First, the $U/t$ dependence, at low temperature (Fig.\ref{opt-dos}(a), both panels).
At small interaction strength $U/t$ the low frequency
Drude weight is large, as we would expect from a metal having finite DOS at Fermi
level, and decreases as we move to higher $U/t$ due to lowering of the DOS at Fermi level.
The Drude weight collapses to zero as we cross to the insulating side when a gap opens
in the DOS, showing a `gapped' response at higher $U/t$.

\begin{figure}[t]
  \centering
  \psfig{file=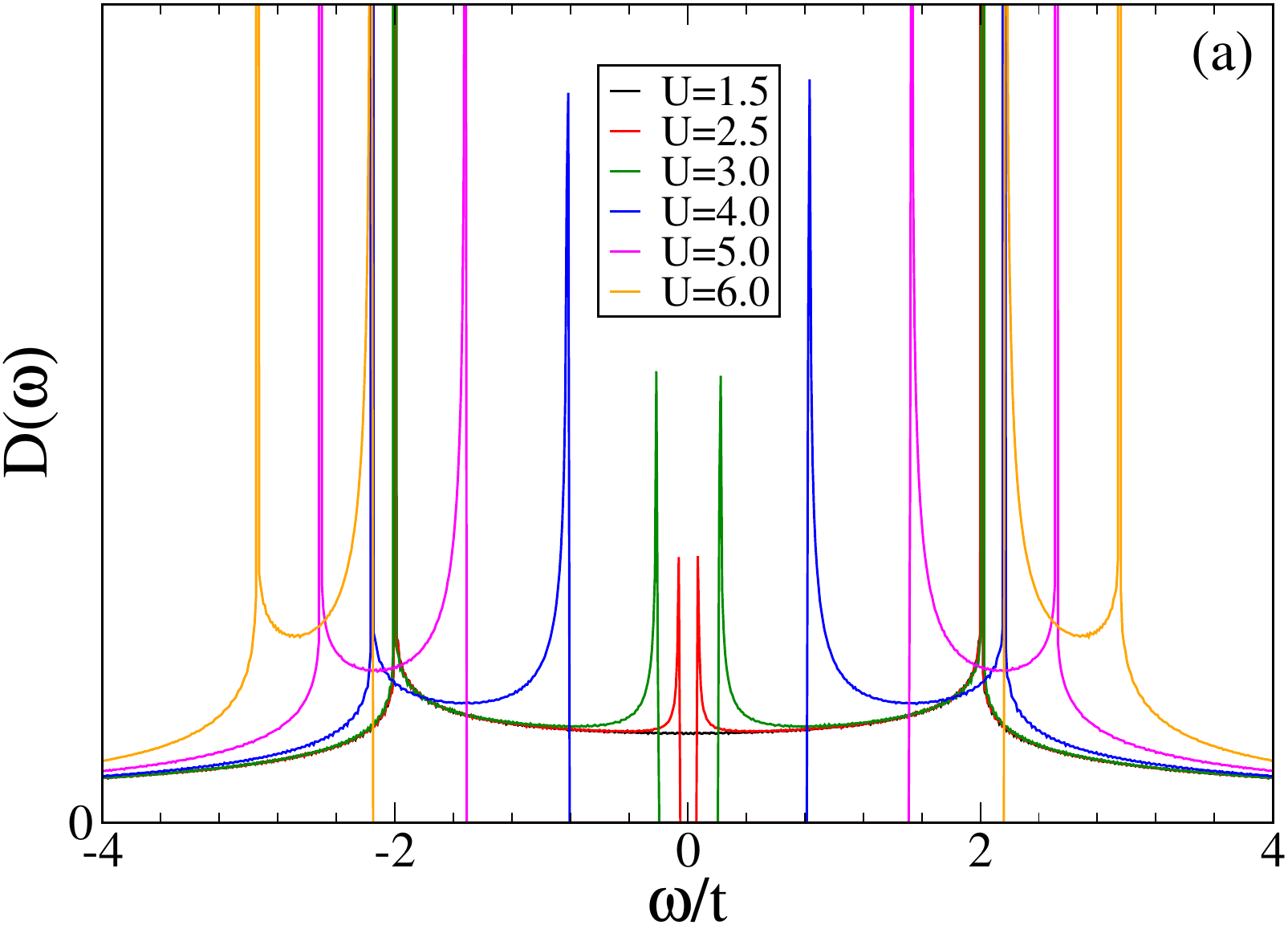,height=5cm, width=8cm}
  \psfig{file=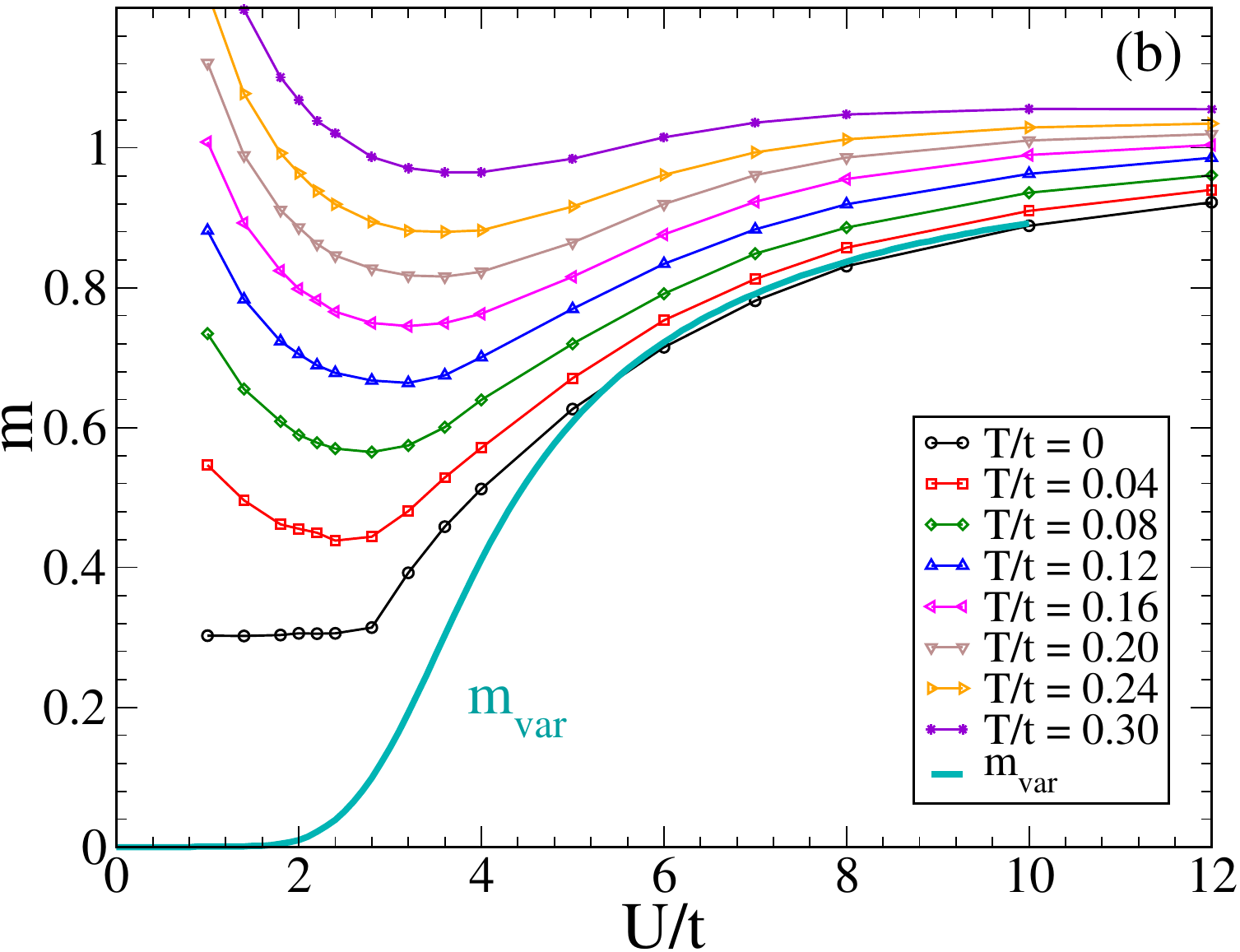,height=5.0cm, width=8.24cm}
  \caption{Colour online:
    (a) The DOS at variationally minimized grounds state at different $U/t$.
    (b) The average magnetization $m$ and $m_{var}$ as function of $U/t$ at different temperature.}
  \label{avgm-dos}
\end{figure}

In the optics, interestingly, we have a `two-peak' behaviour at low temperature.
The low energy peak is the Drude peak, which is strong at weak interaction,
systematically shifts its weight from $\omega=0$ to higher energy, to $\omega \approx mU$,
$m$ being the average magnitude of the auxiliary field ${\bf m}_i$, and is absent
at large $U/t$ in the insulating state.
The second peak, typically prominent only in metallic side, can be attributed
to the TB part of the checkerboard lattice. The TB DOS, shown in the right panel,
has divergence at $\omega/t\approx\pm 2$, which would result in strong response
in optical conductivity around $\omega/t \approx 4$, hence the second peak.
This is what we see in the corresponding optical conductivity panel (compare (c) in both panels)
when the interaction is small, system metallic, and auxiliary field magnitudes
$m_i$ dictating the spectrum are small. With increasing interaction,
the first peak progressively moves from $\omega/t = 0$, to $\omega/t \approx mU$,
while the second peaks stays close to $\omega/t \approx 4$.

The divergence in the DOS, also results in making the finite size effects more
severe in the metallic sides at low temperature, when auxiliary field magnitudes
$m_i$ are small, and the spectrum of the system resembles closer to that of TB system.
This is seen in DOS panel, where the low temperature DOS in metallic side (c) has
large large fluctuations, which become smoother as one increases U
(see panels (d) and (e)), or temperature.

To check the quality of our Monte-Carlo annealing, we also estimated the ground
state of the eq.\ref{eq:hub-hs} using variational minimization. We constructed spiral
configurations ${\bf m}({\bf r}) = m(\cos{\bf q\cdot r}, \sin{\bf q\cdot r}, 0)$
as variational states and minimized the total energy with respect to the magnitude
$m$ and period vector ${\bf q}$. Such periodic states can be easily diagonalized in
Fourier space, as one gets only off diagonal matrix elements connecting
$|{\bf k}\uparrow \rangle$  $\rightarrow$ $|{\bf k+q}\downarrow \rangle$ and back from
the ${\bf m}({\bf r})$ dependent term. We first minimized for both ${\bf q}$ and $m$
over the phase diagram on larger lattice $N = 48\times 48$, and saw that the
${\bf q}$ that minimizes the energy throughout
the interaction was ferromagnetic, i.e., ${\bf q}=(0,0)$, with $U$ dependent
$m = m_{var}$. We then calculated optimal $m_{var}$ for ferromagnetic
phase as function of $U/t$ over even larger lattice $N=2000\times 2000$.
In Fig.\ref{avgm-dos}(a), we show the DOS of the variationally minimized
ferromagnetic state at different $U/t$. In Fig.\ref{avgm-dos}(b), we show
$U/t$ dependence of the average auxiliary field magnitude $m$ calculated
at different temperature from Monte Carlo, and compare it with the variationally
minimized value $m_{var}$. As we see, the MC estimates of average $m$ have non-zero
values at low T and low U.
This is actually consequence of MC annealing, rather than finite size, since one
samples random ${\bf m}_i$ vectors uniformly from inside a sphere of radius, say $m_0$
for Monte Carlo, and one would largely generate vectors with magnitude $m>0$, even if
small, due to zero measure of point $m=0$. Thus the MC picks up $|m| > 0$, even in the
$U<U_c$ side at very low temperature. However, such issues do not occur, for example in
similar simulations of say double exchange model, where the spins have unit magnitude.

At small $U$, when the optimal $m_{var} = 0$, we get the TB ground state, and corresponding
DOS, that is nearly constant at half filling, with singularity at $\pm 2$. Around
$U\approx 2.5$, the optimal $m_{var}$ starts growing (panel (b)), and a gap opens
at half filling, and each singularity splits into two (panel (a)).
%The behaviour of the averaged auxiliary field is $m$
For moderate to strong interactions, $U/t \gtrsim 3$, or $U_c$, the average of
the auxiliary field $m$ is monotonic, i.e., it increases with $U/t$ as well as temperature,
and in large $U/t$ limit, starts to saturate towards atomic limit. The lowest temperature
MC estimate matches with the variational estimate at large interaction.
At lower interaction, however, we notice that (i) the $m(U)$ profile for a given temperature
is non-monotonic, and (ii) the lowest temperature $m$ from MC is higher than the variational one.
The later, results as consequence of the finite size effect, which in combination of large
degeneracy in the spectrum close to and in metallic side, becomes more severe.
The non-monotonicity results from the fact that at very low $U$, when the system's ground
state is $m=0$, finite temperature fluctuations require larger $m$, at smaller $U$, as the
actual fluctuations in the spectrum depend on $Um_i$.

\section{Discussion}
\label{discuss}
% {\bf needs writing, some portion of the above paragraph may be merged here.}
We studied the complimentary scenario of weak to zero super-exchange
limit, to explore the Mott transition only in term of U/t interaction window.
We have done a comprehensive study of the model defined in eq.(\ref{furukawa}),
on checkerboard lattice, which shows strong correlation driven Mott transition with
ferro-orbital order. In term of comparison with real materials, unfortunately,
since majority of the pyrochlore compounds, aren't half filled orbitals
system with weak super-exchange, where the MIT can be seen as purely
Mott phenomenon. There are ferromagnets such as Nd$_2$Mo$_2$O$_7$ and
Sm$_2$Mo$_2$O$_7$\cite{iguchi1}, but these are metals.
We now comment on some limitations, and simplifications
used in our model. It is worth recalling the assumption that the
underlying spin-ordering remains ferro at temperatures well above
orbital ordering temperature $\sim T_{corr}$ shown in figure \ref{fig-phd-sf}.
If that assumption is relaxed, say the spin-ferro ordering temperature $T_c$
becomes comparable to the orbital $T_{corr}$, the core spins, of the model (eq.\ref{DEmodel})
can not be treated as frozen, and the electronic hopping becomes
angle dependent. This would happen when we switch on the super-exchange
interaction $J_{AF}$.
When $J_{AF} \ne 0$, but small, the long range order of the core spins would
still be ferromagnetic, however, with lower $T_c$. For larger $J_{AF}$, the
ordering of the core spins would become antiferromagnetic.
In either case, the electronic hopping would become angle dependent through
spins (see eq.(\ref{hopelement})). One would need to include the core spins,
in the simulations, as the spin fluctuations driven hopping would crucially
impact electronic properties, including transport. As a result, the
effective electronic delocalisation would reduce, pushing the Mott boundary
to lower values, along with possibly reducing the $T_{corr}$ scales.
As mentioned above, for large $J_AF$, more accurate treatment would require
solving the current model with spin dependent hopping, and we plan to
present such a work separately in future.

We considered only the inter-orbital hopping in the current work, which
led to ferro-orbital order. However in reality, the intra-orbital hopping
would also be non-zero. To understand its impact, consider the two limits,
(i) the spin conserving limit, $t_{\alpha \ne \beta} = 0$, the effective exchange between sites         
as seen at half filling through virtual hopping is AF with exchange $\sim t^2/U$ at large U.           
(ii) with spin conserving term being zero, $t_{\alpha=\beta} = 0$, where the
corresponding exchange is ferro due to alternating hopping.
When both hoppings are present and comparable, so are the corresponding competing
exchanges, which would lead to emergence of other long-range orders. Such scenario,
even without the complications of super-exchange, is an intriguing test bed of
frustration in orbital space, and warrants further investigation.

We also wish to comment on phase diagram near the Mott transition, in Fig.\ref{rho} (right),
where we plot the MIT boundary, the $T_{corr}$ curve, and DOS at Fermi level as color.
The Mott transition, usually reflects a windows of pseudo-gap at finite
temperature\cite{tiwari1, nyaya1}, bracketing the MIT curve. Our coloured DOS
with color values between 20\% to up-to $\approx$ 50\% represents a rough estimate of
the pseudo-gap window.

%One could have guessed the ferro ordering from the fact that
%the electron delocalisation happens primarily due to inter-orbital hopping.

%\subsection{Ground state}
% \subsection{Weak coupling limit $t \gg U$: }

\section{Conclusion}
We studied correlation driven orbital Mott transition in two dimensional
pyrochlore lattice using a real space based Monte Carlo approach, and established
finite temperature phase diagram describing the Mott transition, in terms
of MIT boundary, orbital ordering, and a rough estimate of pseudo-gap window.
We also calculated electronic transport, optical conductivity, and thermal
density of states across the Mott transition.

% \begin{figure}[htp]
%   \centering
%   \psfig{file=dos_furukawa.pdf,width=8.0cm}
%   \caption{Colour online:
%     $U/t$ dependence of the DOS $D(\omega)$ at low (a) T/t=0.08, and
%     high (b) T/t=0.30 temperature. Increasing $U/t$ shows systematic
%     opening of depletion of the DOS at Fermi level, and opening of
%     the Mott gap $\sim mU$.
%     In the lower panel we show temperature dependence of $D(\omega)$
%     at three representative (c) $U/t = 2$, (d) $U/t = 4$, (e) $U/t = 6$.
%   }
%   \label{dos}
% \end{figure}

% \acknowledgments
\begin{acknowledgments}
  {R.T. acknowledges the funding provided by the European Research Council
    project QUEST (Project No.307891), and the DJEI/DES/SFI/HEA
    Irish Centre for High-End Computing (ICHEC), projects tcphy108c, tcphy048c
    and Trinity Centre for High Performance Computing (TCHPC) for their
    computational resources and the support of their staff.
    A.S. and R.T. both acknowledge Prof. Pinaki Majumdar for his
    insightful discussions, and encouragements for this work.
  }
\end{acknowledgments}

\end{document}